\documentclass[twocolumn]{aastex631}

\received{2023 July 14}
\revised{2023 October 30}
\accepted{2023 November 2}

\shorttitle{Transit Depth Variability}
\shortauthors{Wang \& Espinoza}

\begin{document}

\title{A Blind Search for Transit Depth Variability with \textit{TESS}}

\author[0000-0003-3092-4418]{Gavin Wang}
\correspondingauthor{Gavin Wang}
\email{gwang59@jhu.edu}
\affiliation{Department of Physics \& Astronomy, Johns Hopkins University, Baltimore, MD 21218, USA}

\author[0000-0001-9513-1449]{Néstor Espinoza}
\affiliation{Space Telescope Science Institute, Baltimore, MD 21218, USA}
\affiliation{Department of Physics \& Astronomy, Johns Hopkins University, Baltimore, MD 21218, USA}

\begin{abstract}

The phenomenon of transit depth variability offers a pathway through which processes such as exoplanet atmospheric activity and orbital dynamics can be studied. In this work we conduct a blind search for transit depth variations among 330 known planets observed by the \textit{Transiting Exoplanet Survey Satellite} (\textit{TESS}) within its first four years of operation. Through an automated periodogram analysis, we identify four targets (KELT-8b, HAT-P-7b, HIP 65Ab, and TrES-3b) which appear to show significant transit depth variability. We find that KELT-8b's transit depth variability likely comes from contaminating flux from a nearby star, while HIP 65Ab and TrES-3b's apparent variability are probable artifacts due to their grazing orbits. HAT-P-7b indicates signs of variability that possibly originate from the planet or its host star. A population-level analysis does not reveal any significant correlation between transit depth variability and host star effective temperature and mass; such correlation could arise if stellar activity was the cause of depth variations via the Transit Light Source effect. Extrapolating our $\sim 1\%$ detection rate to the upcoming \textit{Roman} mission, predicted to yield of order $100,000$ transiting planets, we expect that $\sim 1,000$ of these targets will be found to exhibit significant transit depth variability. 

\end{abstract}

\keywords{Catalogs (205) --- Exoplanets (498) --- Transit photometry (1709)}

\section{Introduction} \label{sec:intro}

The \textit{Transiting Exoplanet Survey Satellite} (\textit{TESS}) is an all-sky transit survey designed to discover transiting planets orbiting the nearest and brightest stars \citep{10.1117/1.JATIS.1.1.014003}. Though its main aim is to discover new transiting planets, the exquisite precision of \textit{TESS}' lightcurves of stars across the sky allows us to extract additional science for already known planets. This data has been used in previous studies to measure transit parameters to higher precision \citep[see, e.g.,][]{Patel_2022, Ivshina_2022}. 

One signature that can be detected in high-precision lightcurves is transit depth variability, in which the transit depth of a planet varies over time. This phenomenon has been observed for a small handful of exoplanets \citep{Biersteker_2017, https://doi.org/10.48550/arxiv.2212.07450} but has received considerably less attention than other, time-varying phenomena such as transit timing and transit duration variations (see e.g., \citealt{10.1093/mnras/stab3483, Trifonov_2023, Boley_2020}; \citealp[for review see][]{Agol2017}). There are several motivations for studying depth variability. Perhaps the one which has recently gained most interest is that of atmospheric activity, in which weather and clouds on a planet's atmosphere cause the opacity and thus apparent size of the planet to change over time \citep{Spiegel_2009, Parmentier_2013, Komacek_2020}. For instance, atmospheric variability has been suggested for Kepler-76b \citep{Jackson_2019}, WASP-12b \citep{Bell_2019}, and HAT-P-7b \citep{Armstrong_2016} through phase curve analyses, although recent evidence from \citet{Lally_2022} may suggest otherwise for the latter. WASP-121b has also been suggested to exhibit atmospheric variability from transit spectroscopy \citep{10.1093/mnras/stab797}. Given the sensitivity of \textit{TESS}, transit depth variability due to such atmospheric activity could be detectable from its data, allowing us to construct a list of planets with possible atmospheric activity. Furthermore, since \textit{TESS} focuses primarily on the nearest and brightest planets, such targets could then easily be followed up by ground- and space-based observing programs. 

An additional motivation for studying depth variations is its links to stellar activity, which may present itself as periodic modulations in the lightcurve through the Transit Light Source effect \citep{Rackham_2018}. Inhomogeneous features such as starspots and faculae contaminate the face of the planet's host star, and as the size and number of these features change over the activity cycle of the host star, so does the amount of contamination in the transits \citep[see, e.g., Section 2 of][]{Morris_2018}. Variability that could be due to starspots has been analyzed for planets including KIC 12557548b, which was found to have a significant transit depth -- rotation modulation signal \citep{Kawahara_2013, 10.1093/mnras/stv297}. However, to our knowledge no blind search for transit depth variability has been conducted as of yet. 

Orbital dynamics, including orbital precession and inclination-eccentricity cycling, also induce changes in an exoplanet's orbit and have been found to cause transit depth variations. A recent example is the case of the warm Jupiter TOI-216b \citep{https://doi.org/10.48550/arxiv.2212.07450}, whose orbit changed from grazing to non-grazing over the course of two years, resulting in a larger planet-to-star radius ratio and a refined inference of the planet's size. Thus, studying transit depth variability also enables us to place constraints on exoplanetary orbital dynamics. Planetary oblateness has been suggested as well to give rise to depth variations \citep{Carter_2010, Biersteker_2017}. 

Measuring transit depth variations using \textit{TESS} data gives us indications regarding the stability of \textit{TESS}, too. In particular, since \textit{TESS} has large 21" pixels and observes the sky by splitting it into different sectors, the varying orientation of \textit{TESS}' cameras and positions of stars over the course of different sectors may result in time-dependent blending. Such effects, if left unaccounted for, directly propagate into changes in the transit depth, and thus the level of consistency of transit depths between sectors informs us of the long-term stability of \textit{TESS} across sectors. 

Finally, while we do not study secondary eclipse depth variability in the present work, we note that such measurements more directly probe temperature and cloud variations in planetary atmospheres. Such variability claims have been made for planets including WASP-12 b \citep{Hooton_2019, von_Essen_2019} and 55 Cnc e \citep{Demory_2016, Tamburo_2018}. We believe that a population-level analysis analogous to the one we conduct here but for secondary eclipse depths is an interesting and exciting study for future work. 

In this work we present a blind search for transit depth variability among a broad sample of known planets observed by \textit{TESS}. In Section \ref{sec:data}, we describe our sample selection, data modeling processes, and statistical metrics used to measure transit depth variability. Section \ref{sec:results} details the results of our analyses, and in Section \ref{sec:conclusion} we discuss the implications and conclusions of our work. 

\section{Data \& Modeling} \label{sec:data}

\subsection{Selection of Targets} \label{subsec:targets}

At the time of our analysis (October 2022), the NASA Exoplanet Archive\footnote{\url{https://exoplanetarchive.ipac.caltech.edu/}} showed that there were more than 5,000 discovered planets, of which 3,944 were known to be transiting. The \textit{TESS} mission had completed 57 sectors at this time, and thus we constructed our sample using all data up to and including Sector 57. We downloaded the available 2-minute \textit{TESS} data for each target. We note that \textit{TESS} data has also been collected in additional cadences throughout the course of its mission, including 30-minute, 10-minute, and 200-second cadences. However, in order to keep our analysis consistent across sectors, we chose to use only the 2-minute cadence mode. We then estimated the signal-to-noise ratio for each planet, which we calculated by dividing the transit depth by the median flux uncertainty of the \textit{TESS} normalized lightcurve, scaled by the square root of the number of in-transit points. We obtained the transit depth for each planet through the Exoplanet Characterization Toolkit \citep{matthew_bourque_2021_4556063} software, which in turn extracts the most recent values reported on the NASA Exoplanet Archive. After eliminating targets that did not reach a signal-to-noise ratio of 5, our sample remained of 478 targets. We then further removed targets which had: 

\begin{itemize}
\item Fewer than five transits, which would be likely impacted by small-number statistics and thus prevent the measurement of transit depth variability.
\item Transit timing variations (TTVs) larger than 0.1 days, as our modeling does not account for TTVs. 

\end{itemize}

\begin{figure}[ht!]
    \includegraphics[width=\columnwidth]{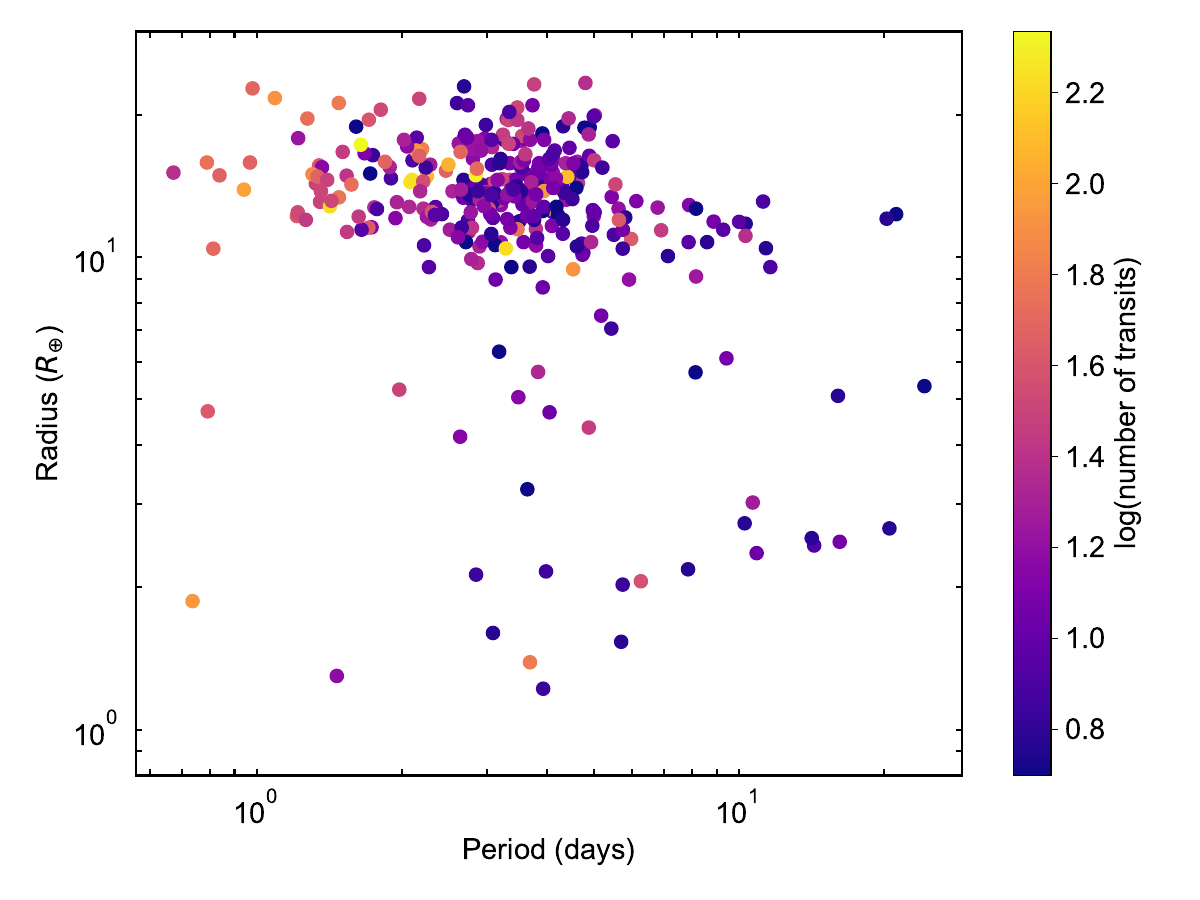}
    \caption{Radius-period distribution for planets in our sample, color-coded by number of transits. The majority of planets in our sample are hot Jupiters with period less than 10 days and radius $\sim 10 - 20 \ R_{\oplus}$. A small number of roughly Earth-size planets are also present.}
    \label{fig:radius-period}
\end{figure}

After these cuts were applied, our sample included 330 targets. Figure \ref{fig:radius-period} shows the radius-period distribution of our sample. Note that, although this work focuses on exoplanets, which indeed comprise the majority of our sample, we did not impose mass or radius constraints to remove brown dwarfs. Therefore, although in this work we will use the terms ``target'' and ``planet'' interchangeably, it should be made clear that not all objects in our sample are, strictly speaking, planets (although they are certainly substellar). 

\subsubsection{Number of Transits}

\begin{figure}[ht!]
    \includegraphics[width=\columnwidth]{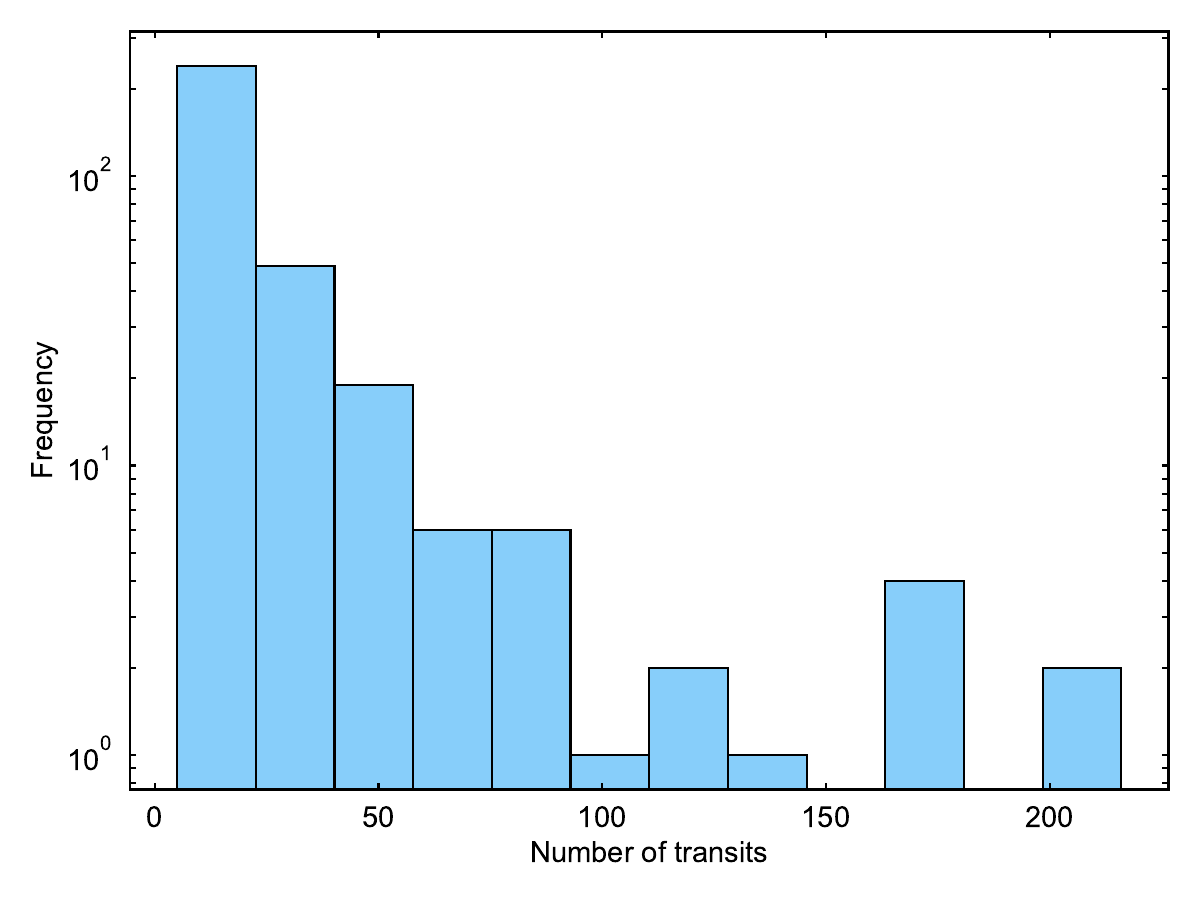}
    \caption{Number of transits for targets in our sample.}
    \label{fig:transits_hist}
\end{figure}

Figure \ref{fig:transits_hist} shows the number of transits for all targets in our sample. In general, the more transits a target has, the more precisely we can measure its transit depth variability. Although the majority of targets in our sample have fewer than 30 transits, there are also a few targets with $\sim 200$ transits for which we can constrain depth variability precisly over long timescales.

\subsection{\textit{TESS} Data Modeling} \label{subsec:modeling}

We largely follow Section 2.2 of \citet{Patel_2022} for our data modeling procedure, using \texttt{juliet} \citep{10.1093/mnras/stz2688} to carry out our analyses. We use the \texttt{batman} \citep{Kreidberg_2015} model and the \texttt{dynesty} \citep{10.1093/mnras/staa278} nested sampler for all of our fits. Before starting our modeling, we queried the Mikulski Archive for Space Telescopes \footnote{\url{https://mast.stsci.edu/portal/Mashup/Clients/Mast/Portal.html}} (MAST) to download all 2-minute cadence Pre-search Data Conditioning Simple Aperture Photometry (PDCSAP) \textit{TESS} lightcurves for targets in our sample. We note that instrumental systematics including momentum dumps and other anomalies have already been corrected for in these lightcurves by the PDCSAP pipeline, and we further remove any values that have a non-zero data quality flag. 

Whereas \citet{Patel_2022} model the data at the (multi-)sector level, here we mainly focus on fitting data at the individual transit level. In particular, for each target, we perform the following two-step analysis. We first use \texttt{celerite}'s \citep{celerite} Quasi-Periodic Gaussian Process (GP) kernel to empirically model the out-of-transit noise, which includes effects such as starspot modulation. For our purposes we define out-of-transit data as data that are captured more than one full transit duration away from the nearest transit center, where we obtain the transit duration, $P$, and $t_0$ from the Exoplanet Characterization Toolkit. We use wide priors for all Gaussian Process hyperparameters (see example priors and posteriors in Table \ref{tab:priors_posteriors}). Secondly, we fit individual transits as informed by our Gaussian Process model, which we accomplish by setting the priors to the posteriors obtained from the out-of-transit fit. Since our goal is to measure transit depth variability across single transits, the fit of one transit should be independent of the others and thus we fit the data one transit at a time. Maintaining consistency with our previous definition, here each transit is fit using data within one transit duration of its transit center. We assign the depth, impact parameter, and quadratic limb darkening coefficients uniform priors between 0 and 1, and the scaled semi-major axis a log-uniform prior from 1 to 100 (see ``In-transit fit'' section of Table \ref{tab:priors_posteriors}). The eccentricity and argument of periastron are fixed to values obtained from the literature, and defaulted to 0 and $90^{\circ}$ respectively if no literature values are available. Although our method is computationally more expensive than a direct fit to the individual transits, it allows us to preserve minute differences between transits. Finally, we visually search through individual transit depths for all targets and remove outliers which were more than $3 \sigma$ away from the surrounding data points, or had an errorbar that was more than $3$ times larger than the median errorbar. 

Though the primary focus of this work is indeed transit depth variability on the single-transit level, we determined that it would also be insightful to conduct sector-level fits for the planets in our sample. These were carried out in an analogous way to the individual transit fits in that we use the same definition of priors as above, with the only difference being that we do not divide the data into segments and instead fit the entire lightcurve for each sector. Such fits inform us of the instrumental stability of \textit{TESS} (see Section \ref{subsec:sigma_results}), in addition to providing us with precise transit parameters that we later use to create noiseless simulations of transits (see Section \ref{subsubsec:simulate}). 

\begin{table*}
    \caption{Sample priors and posteriors for fits of CoRoT-2 b.} \label{tab:priors_posteriors}
    \begin{tabular}{lccr}
        \hline \hline
        Parameter    &Prior    &Posterior  &Description  \\\hline
        \it{Out-of-transit fit} & & &\\
        $m_{dilution}$    &Fixed, $1.0$    &$\ldots$    &Dilution factor\\
        $m_{flux}$    &$\mathcal{N}^{\text{a}} (0.0, 0.1)$    &$0.0029^{+0.014}_{-0.013}$    &Relative flux offset (normalized)\\
        $\sigma_w$    &$\mathcal{J}^{\text{b}} (0.1, 10000.0)$ &$1075^{+77}_{-92}$ &Jitter term (ppm)\\
        $GP_B$    &$\mathcal{J} (10^{-5}, 1000.0)$  &$0.0004^{+0.00093}_{-0.00022}$ &Amplitude of GP (ppm)\\
        $GP_C$    &$\mathcal{J} (10^{-5}, 10000.0)$  & $0.0032^{+0.14}_{-0.0031}$ &GP amplitude additive factor\\
        $GP_L$    &$\mathcal{J} (0.001, 1000.0)$  &$30^{+69}_{-16}$ &Length-scale for GP\\
        $GP_{Prot}$    &$\mathcal{J} (1.0, 100.0)$  &$4.26^{+0.16}_{-0.17}$ &Quasi-periodic GP period\\
        
        \hline
        \it{In-transit fit}\\
        $m_{flux}$    & $*$    &$-0.005^{+0.016}_{-0.017}$    &\\
        $\sigma_{w}$    & $*$ &$1056^{+85}_{-78}$ &\\
        $GP_B$    & $*$ &$0.0012^{+0.0011}_{-0.0007}$ &\\
        $GP_C$    & $*$ & $1.3^{+1.5}_{-0.9}$ &\\
        $GP_L$    & $*$  &$40^{+46}_{-25}$ &\\
        $GP_{Prot}$    & $*$  &$4.26^{+0.18}_{-0.17}$ &\\

        $P$       &$\mathcal{N} (1.7429935, 10^{-6})$  & $1.74299342^{+0.00000095}_{-0.00000091}$ & Orbital period (days)\\
        $t_{0}$      &$\mathcal{N} (2459771.54,0.1)$    & $2459773.29457^{+0.00053}_{-0.00051}$ & Time of transit center ($\text{BJD}_{\text{TDB}}$)\\
        $R_{p}/R_{*}$       &$\mathcal{U} (0.0,1.0)$           & $0.1668^{+0.0057}_{-0.0048}$ & Planet-to-star radius ratio\\
        $b$       &$\mathcal{U}^{\text{c}} (0.0,1.0)$           & $0.24^{+0.15}_{-0.14}$  & Impact parameter\\
        $u_1$          &$\mathcal{U} (0.0,1.0)$           & $0.33^{+0.39}_{-0.24}$ & First limb-darkening coefficient\\
        $u_2$          &$\mathcal{U} (0.0,1.0)$           & $0.17^{+0.31}_{-0.13}$ & Second limb-darkening coefficient\\
        $e$       &Fixed, $0.0143$                   &$\ldots$        & Eccentricity\\
        $\omega$  &Fixed, $102.0$                    &$\ldots$        & Argument of periastron ($^{\circ}$)\\
        $\alpha/R_{*}$  &$\mathcal{J} (1.0, 100.0)$        & $6.43^{+0.19}_{-0.26}$  & Scaled semi-major axis\\
        \hline
    \end{tabular}
    \begin{itemize}
        \item[] $*$ Corresponding posterior for out-of-transit fit
        \item[] $^{\text{a}}$ $\mathcal{N} (a, b)$ represents a Normal prior with median $a$ and standard deviation $b$
        \item[] $^{\text{b}}$ $\mathcal{J} (a, b)$ represents a Jeffreys prior between $a$ and $b$
        \item[] $^{\text{c}}$ $\mathcal{U} (a, b)$ represents a uniform prior between $a$ and $b$
    \end{itemize}
\end{table*}

\subsection{Periodicity Search} \label{subsec:periodogram}

Transit depth variability may emerge in a periodic fashion, whether that be through stellar activity, atmospheric activity, or orbital dynamics. Thus, to identify such periodic signals, we perform a uniform search of periodicity for all targets in our sample using the Lomb-Scargle periodogram \citep{Lomb_1976, Scargle_1982} as implemented through \texttt{astropy} \citep{astropy:2013, astropy:2018, astropy:2022}. Variations in the transit depth with frequencies above the Nyquist frequency or below the maximum continuous time span of the data cannot be reliably detected. Thus, these form the upper and lower bounds for our frequency space, which we uniformly sample with a density corresponding to the minimum resolvable frequency spacing. 

To determine the significance of any peaks detected in the periodogram, we simulate the transit depths with Gaussian white noise and calculate the corresponding 10\%, 1\%, and 0.1\% false-alarm probabilities. Specifically, we assume a fixed transit depth of the system, which we take as the median of the individual transit depths, and set the standard deviation of the Gaussian noise distribution to the median transit depth uncertainty. If a peak for the true data's power spectral density has a false-alarm probability of less than 0.1\%, we consider the peak to be significant. Since there are known complications with claiming significance levels in Lomb-Scargle periodograms \citep{Zechmeister_2008}, for these targets we conduct a second round of vetting by phasing the transit depths to the highest peak and performing a sinusoidal fit to the phased data. This allows us to determine whether the amplitude is consistent with 0 for each target. If the corresponding amplitude for a target is more than $3\sigma$ away from 0, we consider it to be a target of interest and worthy of closer scrutiny. Our results from these searches are discussed in Section \ref{subsec:periodogram_results}. 

\subsection{Population Analysis of Transit Depth Variability} \label{subsec:stellar}

The second objective of this work is to search for correlations between stellar type and transit depth variability on a population level. Achieving this requires us to develop a new metric. Each target has varying baseline noises and thus we cannot directly compare the standard deviations of their transit depths (as calculated in Section \ref{subsec:absresults}). In order to objectively compare depth variations across different planets, the metric that we use must be independent of: 

\begin{itemize}
\item Number of transits, because this is not an intrinsic astrophysical property of the system and is instead a consequence of orbital period and \textit{TESS}' observing baseline.
\item Transit detection signal-to-noise ratio, as this depends on factors including apparent magnitude of the host star and transit depth, which we assume to not be related to true transit depth variability but rather only our ability to observe and measure it. 
\end{itemize}

We next describe our process of simulating and co-adding transits that we use to calculate a metric that satisfies our criteria. 

\subsubsection{Simulating \& Co-Adding Transits} \label{subsubsec:simulate}

\begin{figure*}[ht]
    \centering
    \includegraphics[width=\textwidth]{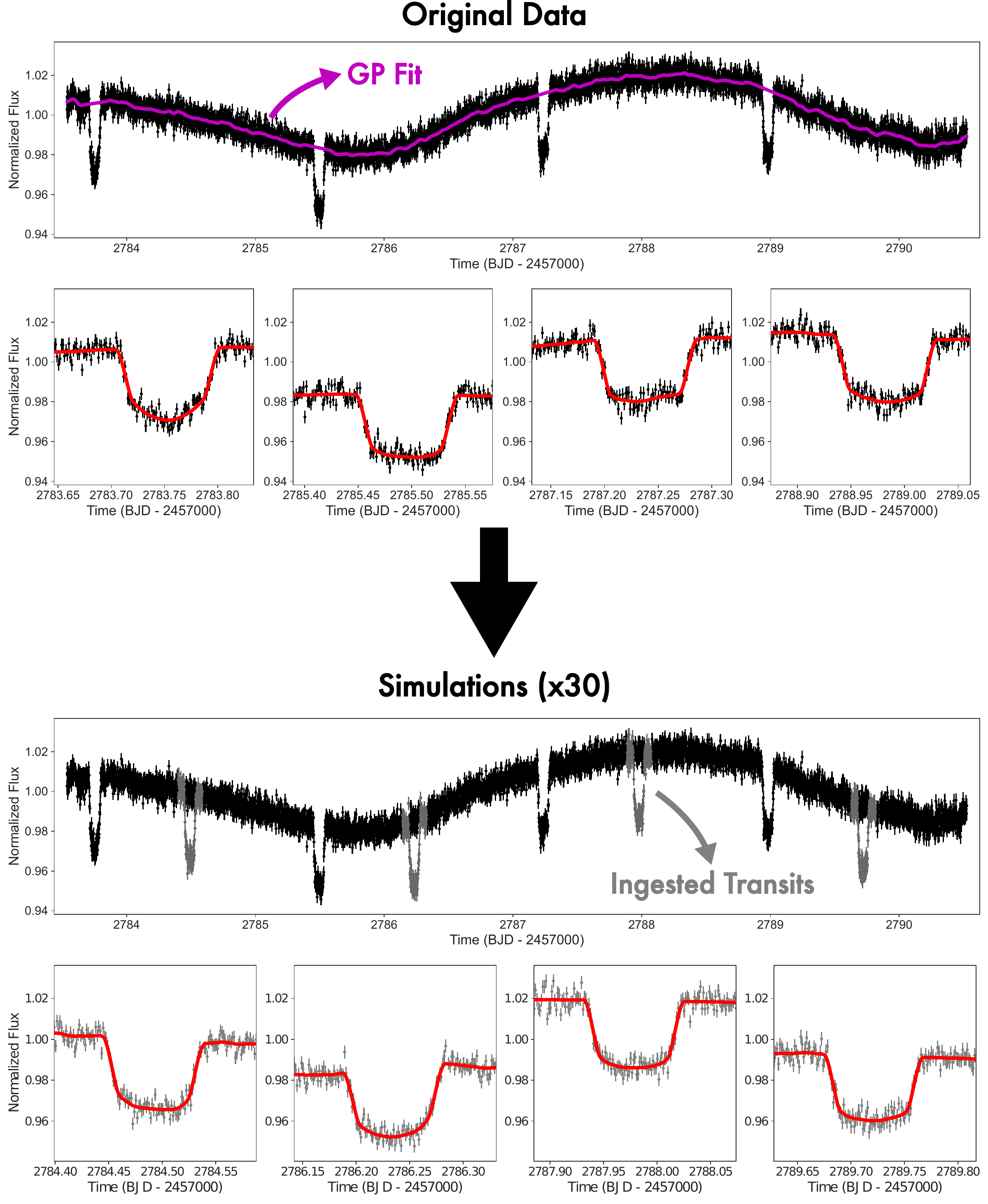}
    \caption{Schematic showing our procedure for calculating the nested ratio.}
    \label{fig:procedure}
\end{figure*}

In broad terms, our approach is to compare variations in the true transit depths against simulated transits. We use the \texttt{batman} model when simulating our transits for consistency with our fits (see Section \ref{subsec:modeling}). For each target, we first extract the posterior parameters from our sector-level fits. These are then input into \texttt{batman}, which returns noiseless, simulated transits with the same 2-minute cadence as the \textit{TESS} lightcurves. It is important to note that all transits within the same sector are modeled using the same set of parameters and thus identical. A consequence of this, however, is that injected transits are slightly different for each sector and this method is thus insensitive to changes occurring over timescales of $>1$ sector. Since we want to explore the effects of stellar type on transit depths, we inject these transits into the out-of-transit baseline of the lightcurve. Specifically, this is done by multiplying point-by-point the original lightcurve flux values by the simulated data. The injection site is chosen to be a time offset from the original transits, which reproduces a full set of transits, each of which are offset by the same time length from the original transits. Finally, we repeat this process for a total of 30 sets per target, with each set corresponding to incremental offsets such that the entire lightcurve baseline is sampled. This process is shown in Figure \ref{fig:procedure}. 

\begin{figure}[ht!]
    \centering
    \includegraphics[width=\columnwidth]{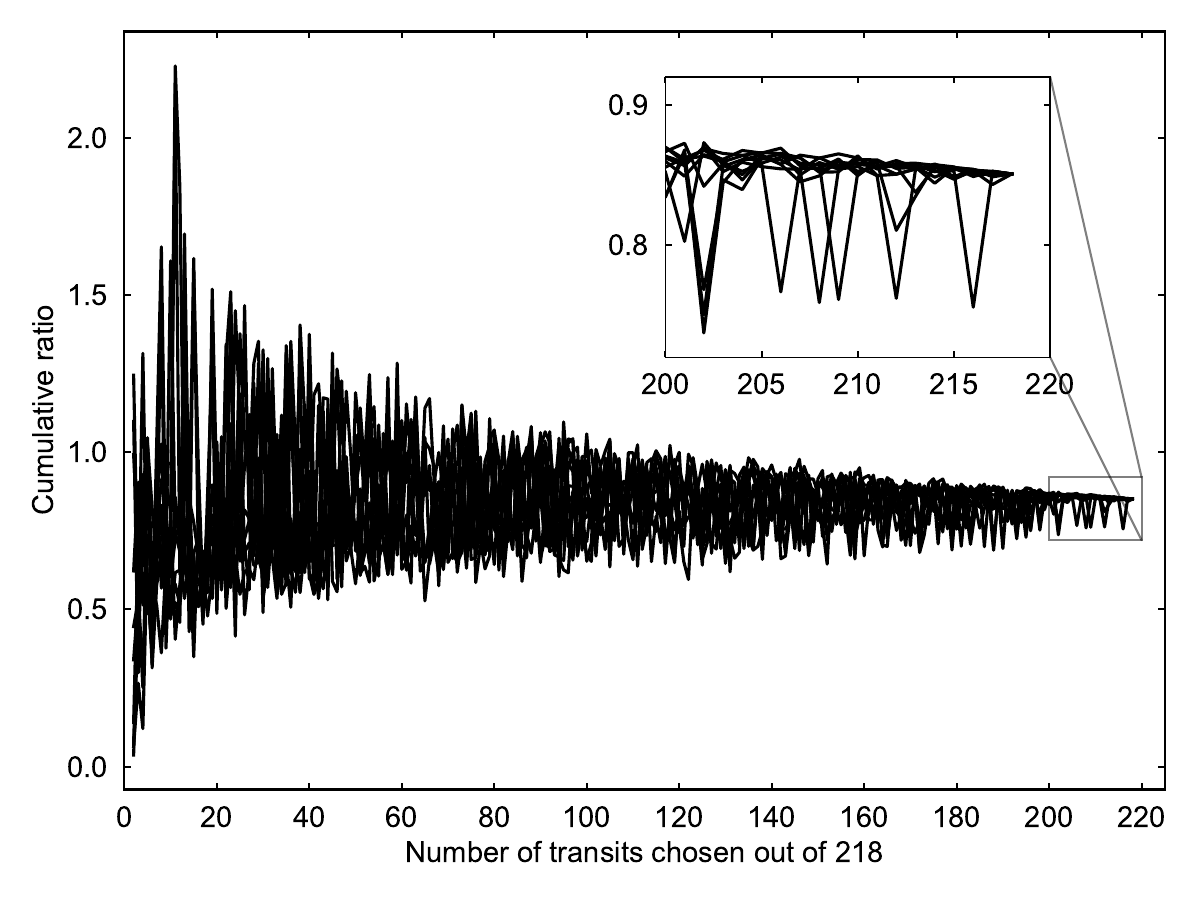}
    \caption{Ratios for randomly-selected samples of different sizes for Qatar-10b. Ratios are calculated starting from 2 randomly-chosen transits until all 218 transits for the target are included. Even omitting just a small number of transits can result in a large change in the ratio, demonstrating that the F-ratio is not immune to changes in the number of transits.}
    \label{fig:cumulative_ratio}
\end{figure}

In reality, the process of injections is complicated by the fact that there are data gaps in the \textit{TESS} data. Due to these gaps, it is often the case that a set of injections is unable to contain as many transits as the true lightcurve. Our analyses reveal that the F-ratio, which is the standard deviation of the transit depths divided by the median uncertainty of the transit depths, can exhibit large changes even with small ($\sim 1\%$) changes in the number of transits used to calculate the ratio, as shown in Figure \ref{fig:cumulative_ratio}. Thus, we enforce the constraint that the number of transits in a set must be exactly equal to that in the true lightcurve to avoid introducing additional sources of error. As a general rule of thumb, we accept targets which have $\geq 9$ successful injections out of 30. For targets whose success rates are low ($< 30\%$), we perform a second round of injections, increasing the number of injections to 100 and maintaining inclusion of targets with $9$ or more transits. In this way, we are able to incorporate a larger percentage of our original sample into our population-wide analysis. 

\subsubsection{Nested Ratio from Simulated Transits} \label{subsubsec:nested}

For each target, we have between 9 and 30 sets of simulated transits, and we calculate the F-ratio for each set. We also repeat this calculation for the true transit depths to obtain $F_{t}$. Finally, we normalize $F_{t}$ by dividing it by the distribution of simulated F-ratios. Hereafter in this text we will call this ratio of the F-ratios the ``nested ratio,'' which is the metric that we will use to search for population-level trends: 

\begin{equation}
    N \equiv \frac{F_{t}}{{F}_{i}}
\end{equation}

where ${F}_{i}$ is the median of the injected F-ratios. 
It should be noted that F-ratios for neither the true nor simulated transits have associated uncertainties. However, we can make use of the fact that we have a distribution of simulated F-ratios to measure the uncertainty of $N$. For a target which has $n$ injections and thus $n$ simulated ratios, we randomly choose two-thirds of those ratios, whose median we will call $F_{2/3}$. We can calculate one value of $N$ by taking $N = F_{t} / F_{2/3}$. For each planet, we repeat this process $100$ times, resulting in $100$ random draws of $N$. Our nested ratio is then the median of this distribution, with associated uncertainty $\sigma_{N}$ as the standard deviation of the distribution.

\section{Results} \label{sec:results}

\subsection{Absolute Transit Depth Variability} \label{subsec:absresults}

Using the steps outlined in Section \ref{subsec:modeling}, we obtain a list of transit depths $\delta$ and uncertainties $\sigma_{\delta}$ for each target. In order to empirically measure the variability in the transit depth for each target, we calculate the standard deviation $\sigma$ in the transit depth for each target in units of parts per million (ppm). We call this value the absolute transit depth variability. Note that this is different from the per-point uncertainty $\sigma_{\delta}$ in the transit depth. However, if there is no inherent, excess variability in the transit depth, these two values should converge. 

\begin{figure}
    \centering
    \includegraphics[width=\columnwidth]{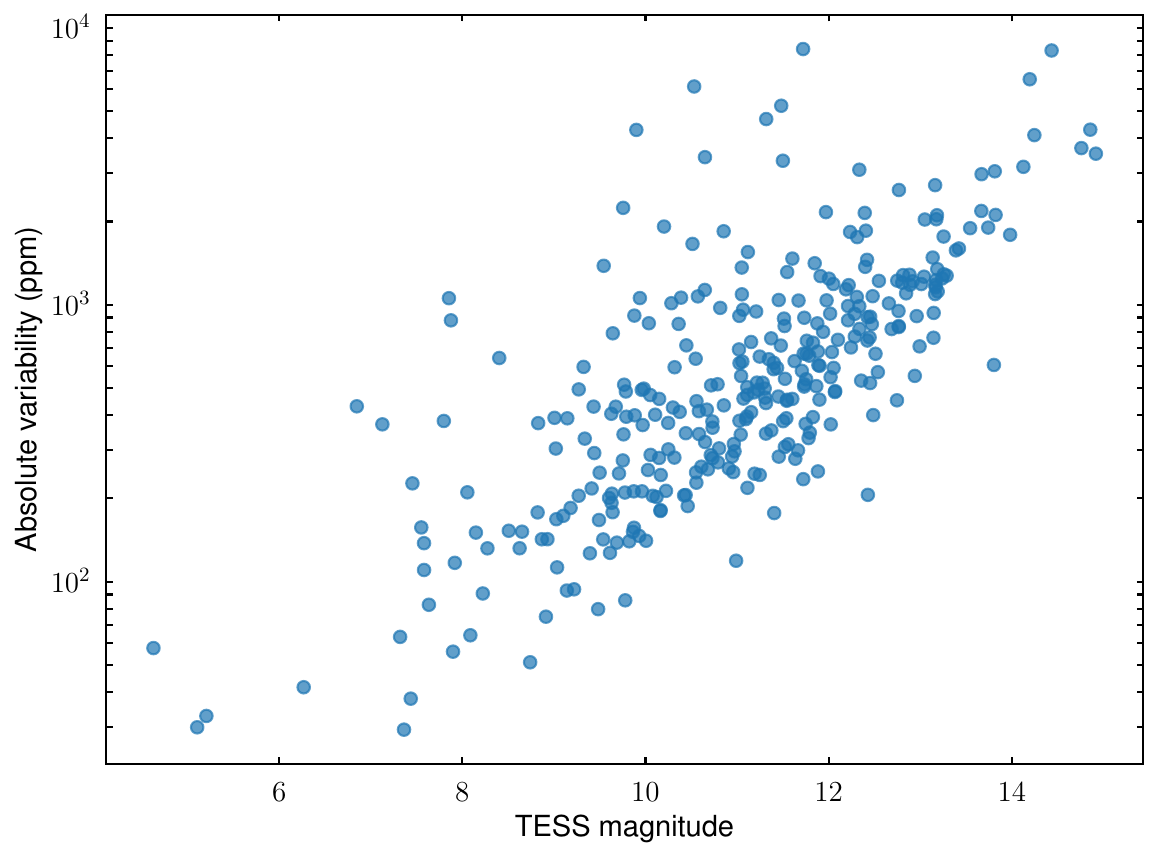}
    \caption{Scatter plot of absolute variability vs TESS magnitude.}
    \label{fig:absvsmag}
\end{figure}

The results from our calculations are listed in Table \ref{tab:mainresults},  with absolute variability typically lying on the order of $1$ part per thousand, comparable to the uncertainties in the transit depths. Figure \ref{fig:absvsmag} plots the absolute variability against \textit{TESS} magnitude for all targets in our sample. We note that the data exhibit a general trend of higher variability for fainter targets that aligns with expectations, albeit with a moderate spread. No planets deviate significantly from this trend.

It is important to note that, although this is an absolute (i.e., measured in parts per million) standard deviation which places absolute constraints on the amount by which the transit depth varies, it is only specific to the data collected by \textit{TESS}. Different instruments and different observing schemes could result in different absolute constraints, but \textit{TESS}' high precision nevertheless makes these values good baseline estimates for future studies.

\subsection{Targets with Sector Discrepancies} \label{subsec:sigma_results}

Blending caused by \textit{TESS}' large pixels is accounted for in the PDCSAP lightcurves through a dilution factor \citep{Stumpe_2012}, calculated based on positions of nearby stars from the \textit{TESS} Input Catalog. However, the proper motion of stars and the orientation of the instrument's four cameras change the pixel positioning of stars over the course of different sectors, which may result in changes in the dilution that are not perfectly accounted for. To measure the instrumental stability of \textit{TESS} across sectors, we conduct an additional search for variability on the sector-level. Specifically, we jointly fit all transits for each sector, and setting the out-of-transit Gaussian Process posteriors as the priors for these sector-level fits. We then calculate the sigma-discrepancy for all pairs of sectors and identify those targets which have a minimum of $3 \sigma$ transit depth discrepancy between sectors. 

We find 3 such targets (TrES-1b, WASP-100b, WASP-183b) out of our sample of 330. We note that finding $\sim 1\% $ of targets with discrepancies above $3 \sigma$ is quite reasonable within a large sample, suggesting that there are no excess biases with \textit{TESS}. Nevertheless, we examine each of these cases in further detail. 

\begin{itemize}
    \item TrES-1b was observed by \textit{TESS} in Sectors 14, 40, 41, 53 and 54. For Sectors 14 and 53, our fits obtained $R_{p}/R_{*}$ values of $0.1395 \pm 0.0014$ and $0.1308^{+0.0023}_{-0.0020}$, respectively. These two values are $3.17 \sigma$ apart, while all other pairs of sectors were consistent to within $3 \sigma$ of each other. We believe this is in part due to an anomalous transit in Sector 53 which falls in a data gap due to a momentum dump and which has a depth that is roughly only half of the other transits, thus biasing the fit towards a lower $R_{p}/R_{*}$. Although we had manually removed all outlier individual transit depths (see Section \ref{subsec:modeling}), we did not remove those discrepant transit depths from the sector-level fits. To confirm our hypothesis, we repeated the sector fit for Sector 53 with the anomalous transit removed, which returns $R_p / R_{*} = 0.1322^{+0.0021}_{-0.0019}$. This is now consistent to within $2.89$ sigma with the depth from Sector 14. Although nominally consistent, we believe that this target is interesting and warrants further scrutiny. 
    \item WASP-100b falls in the \textit{TESS} Continuous Viewing Zone and was observed in 20 sectors. The planet-to-star radius ratios for Sectors 2 and 12 are $0.0799^{+0.0014}_{-0.0010}$ and $0.0856^{+0.0009}_{-0.0010}$, discrepant at $3.30 \sigma$. Sector 2 is also discrepant with Sector 10 ($R_{p} / R_{*} = 0.0847 \pm 0.0007$) by $3.16 \sigma$, and there were no other discrepant pairs of sectors. We note that the PDCSAP dilution factor for Sector 12 was 0.88, significantly smaller than the $0.92 - 0.93$ values for other sectors both before and after it. The dilution factor is absent in the FITS headers for Sector 10, and we hypothesize that variations in the dilution factor caused the observed discrepancy. An analysis of the typical variation in the \textit{TESS} dilution factor, however, reveals that $20$ of $107$ arbitrarily chosen planets in our sample exhibit changes in the dilution factor that are larger than $0.04$, and yet do not display significant sector-level variations. Thus, we believe that WASP-100b is an interesting target to further monitor as well.
    \item WASP-183b was observed in Sectors 35, 45, and 46. The best-fit depths for Sectors 45 and 46 are respectively $0.1229^{+0.0061}_{-0.0042}$ and $0.1500^{+0.0053}_{-0.0057}$ and discrepant by $3.25 \sigma$. We attribute this to the high impact parameter of the target, resulting in a degeneracy in the transit depth. This is supported by the V-shaped transits, in addition to the impact parameter posterior values of $b=0.30^{+0.18}_{-0.14}$ (Sector 45) and $b=0.75^{+0.05}_{-0.06}$ (Sector 46) for these two sectors, which differ by $2.4 \sigma$. We additionally note that our transit-by-transit fits do not show such large changes in $b$, and upon constraining the limb darkening coefficients around theoretical values, the discrepancy between the two sectors disappears. However, we still cannot rule out variability such as orbital dynamics or precession on timescales of $\sim 1$ sector ($27$ days) resulting in a variable inclination and thus impact parameter. 
\end{itemize}

Collectively, these results suggest that one must take caution when analyzing targets with transits that fall near a data gap and those with high impact parameters. Additionally, this analysis demonstrates that \textit{TESS} is largely able to accurately account for and correct variations in dilution for a broad sample of targets. 

\subsection{Periodogram Search} \label{subsec:periodogram_results}

We repeat the procedure outlined in Section \ref{subsec:periodogram} for all targets in our sample. In most cases, we do not find any statistically significant peaks above what is expected from Gaussian white noise. However, for 4 targets (HAT-P-7b, KELT-8b, TrES-3b, HIP 65Ab) we discover peaks that fall above 99.9\% of the points obtained through bootstrap resampling, corresponding to false-alarm probabilities of less than 0.1\%. 

\begin{figure*}[ht!]
    \centering
    \includegraphics[width=0.92\textwidth]{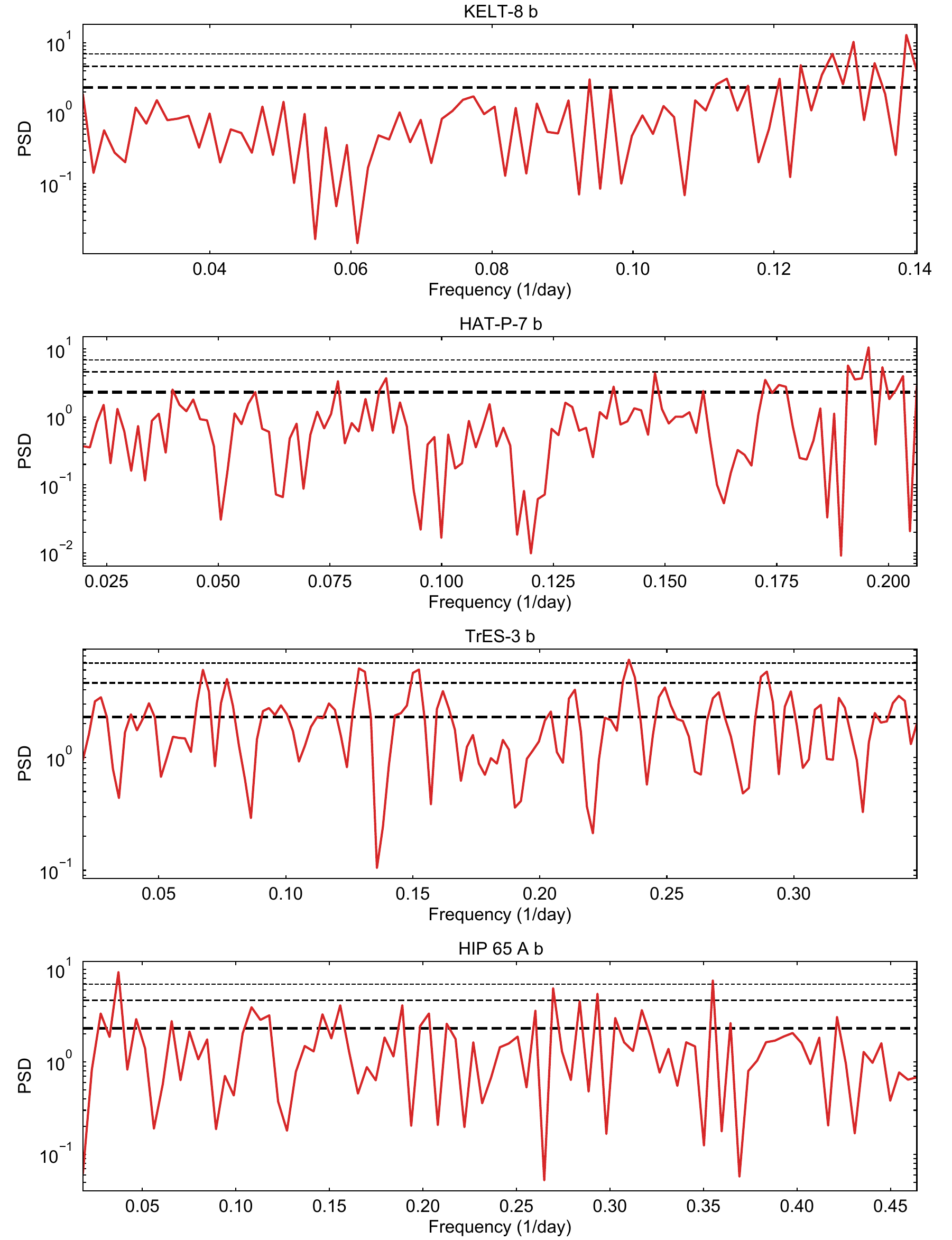}
    \caption{Periodograms for KELT-8b, HAT-P-7b, TrES-3b, and HIP 65 Ab from top to bottom. In each panel, horizontal dashed lines indicate 10\%, 1\%, and 0.1\% false-alarm probabilities (from bottom to top). All targets have peaks with false-alarm probabilities of less than 0.1\%.}
    \label{fig:periodograms}
    \clearpage
\end{figure*}

\begin{figure*}[ht!]
    \centering
    \includegraphics[width=\textwidth]{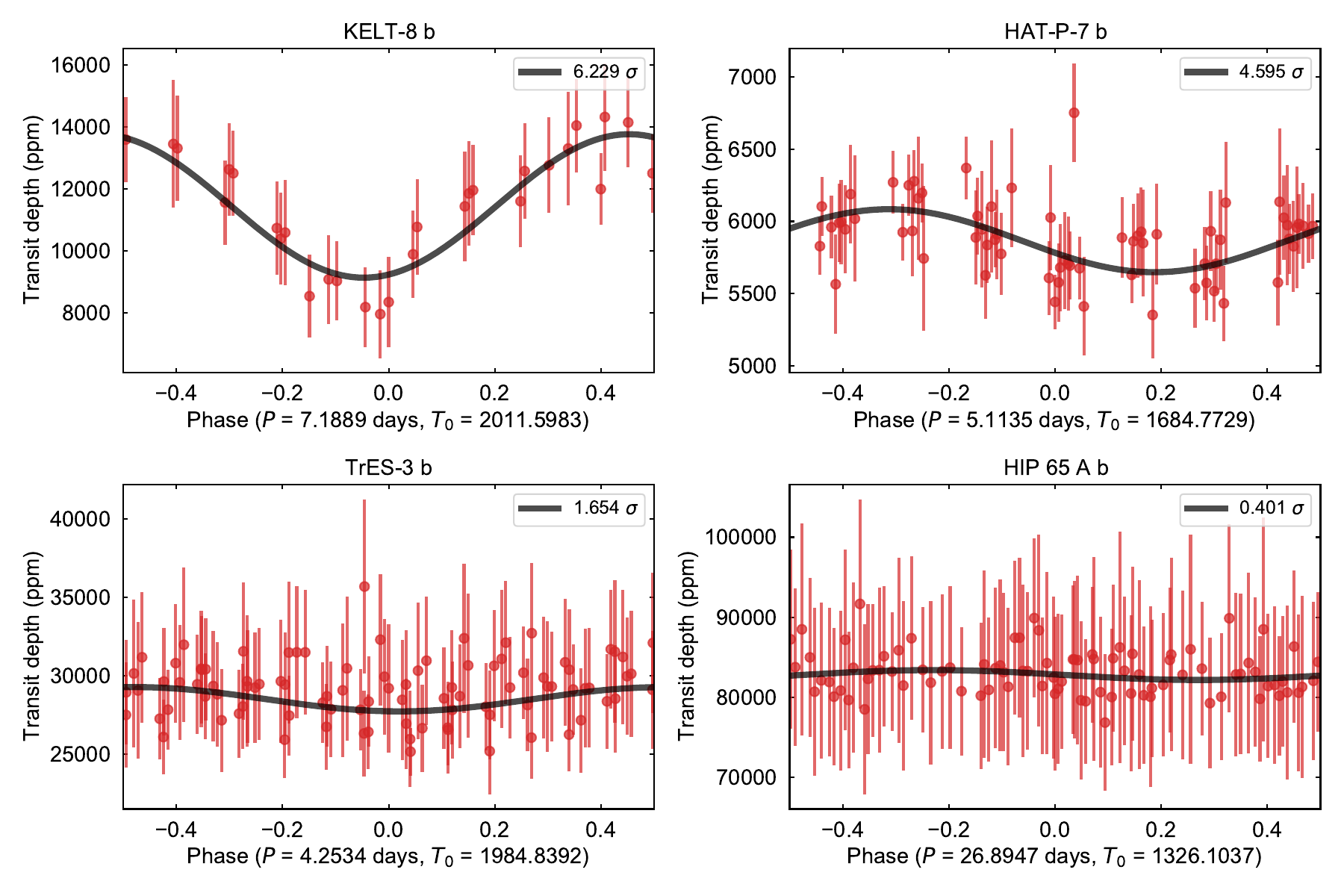}
    \caption{Phased transit depths over time for KELT-8b, HAT-P-7b, TrES-3b, and HIP 65 Ab. The transits depths for TrES-3b and HIP 65 Ab are from a second fit using an impact parameter fixed to the literature value. The best-fit sinusoidal models are overplotted in black and labeled with the amplitude significance.}
    \label{fig:phased}
\end{figure*}

Continuing the vetting, for each of these 4 targets we calculated the significance of the amplitudes, finding that all 4 have amplitudes of more than $3\sigma$ from zero (see Table \ref{tab:sigmas}). However, the transits for TrES-3b and HIP 65Ab initially had unusually large uncertainties, which we determined to be due to their high impact parameters and our fitting choices (see Sections \ref{subsubsec:tres-3b} and \ref{subsubsec:hip-65ab} for a detailed investigation). Moving forward we will only present the results from the re-analysis which fixes the impact parameter to literature values. For each target, the transit depths are phased to the highest frequency peak found in Figure \ref{fig:periodograms} and plotted in Figure \ref{fig:phased}. We discuss each of these targets below. 

\begin{table}[ht!]
    \caption{Amplitude sigmas for each of the 4 targets. TrES-3b and HIP 65Ab have values from a re-anaylsis using a fixed impact parameter, which greatly reduces the significance of their transit depth variability signals.} \label{tab:sigmas}
    \centering
    \begin{tabular}{lcc}
        \hline \hline
        Planet    &Amplitude Significance ($\sigma$) &Re-Analysis ($\sigma$)\\ \hline
        KELT-8b &6.229 &$\ldots$ \\
        HAT-P-7b &4.595 &$\ldots$ \\
        TrES-3b &3.811 &1.654\\
        HIP 65Ab &4.313 &0.401\\
        \hline
    \end{tabular}
\end{table}

\subsubsection{KELT-8b} \label{subsubsec:kelt-8b}

KELT-8b, also known as TOI-2132.01 and TIC 358516596.01, is a hot Jupiter with a 3.2-day period \citep{Fulton_2015}. This target was observed by \textit{TESS} in Sectors 26, 40, 53, and 54. \citet{Fulton_2015} measured a best-fit transit depth of $13100 \pm 600$ ppm, whereas our fits find clear variability in the depth ranging from values as extreme as 8000 to 14000 ppm with amplitude $2320 \pm 370$ ppm, or $6.2 \sigma$ (see Figure \ref{fig:phased}). Furthermore, as shown in Figure \ref{fig:oot}, we find strong modulations in the lightcurve with period $0.197$ days and amplitude $3145 \pm 9$ ppm. 

\textit{Gaia} epoch photometry and \textit{TESS} centroid offsets reveal, however, that the modulations in the lightcurve are due to a blended variable star (TIC 358516610, Gaia DR3 4534144923690486144, $T = 13.967$) 25" from the target star ($T = 10.203$). After accounting for blending and the brightness ratio of the two stars, the variations match what is observed in KELT-8's lightcurve in both period and amplitude. Thus, we conclude that this is a false positive scenario, and that the true transit depth for KELT-8b likely does not vary in the absence of the nearby contaminating source. 

Nevertheless, this candidate serves as a reminder for the need to be cautious of blending scenarios when claiming precise transit signatures with \textit{TESS} data. Furthermore, the fact that KELT-8b experiences transit depth variability with a period of 7.2 days, unrelated to both its orbital period of 3.2 days and the contaminating source's period of 0.2 days demonstrates the complex effects that photometric activity can have on transit depths. 

\subsubsection{HAT-P-7b} \label{subsubsec:hat-p-7b}

HAT-P-7b (Kepler-2b, TOI-1265.01, and TIC 424865156.01) was observed in \textit{TESS} Sectors 14, 15, 40, 41, 54, and 55. This is a massive hot Jupiter with a 2.2-day period and high planetary irradiance \citep{Pal_2008}. We find the median transit depth to be $5900 \pm 250$ ppm, consistent with \citet{Pal_2008}'s value of $5800 \pm 150$ ppm, but we also find that the transit depth varies on a 5.1-day period with amplitude $218 \pm 47$ ppm ($4.6 \sigma$). 

The strongest photometric variability detected in the out-of-transit portion of the lightcurve has variation amplitude $93 \pm 5$ ppm and period 6.53 days. There are a few close sources nearby to HAT-P-7. Thus, it is possible that, as with the case of KELT-8, the detected photometric variability is actually due to activity from a nearby star. However, we are unable to make definitive conclusions without further data, as starspots and faculae on the host star HAT-P-7 could also be capable of reproducing the same observed variability. 

We note that HAT-P-7b has actually had an interesting history of atmospheric variability claims. Using \textit{Kepler} data, \citet{Armstrong_2016} detected variations in its phase curve, with the position of the peak shifting longitudinally between orbits. These variations were initially attributed to atmospheric variability. More recently, however, \citet{Lally_2022} suggested that the observed variations could instead be caused by stellar activity. Nonetheless, no previous work known to the authors has reported on transit depth variations for this target, and thus our work may provide a new perspective from which follow-up studies can continue to examine the nature of HAT-P-7b and its atmosphere (with, e.g., transmission spectroscopy). 

\subsubsection{TrES-3b} \label{subsubsec:tres-3b}

TrES-3b \citep{Donovan_2007}, also known as TOI-2126.01 and TIC 116264089.01, is a hot Jupiter with a short orbital period of 1.3 days. This target was observed by \textit{TESS} in Sectors 25, 26, 40, 52, and 53. The discovery paper found its transit depth to be $27600 \pm 800$ ppm. Although this is consistent with the median transit depth of $29800 \pm 8300$ ppm from our initial fits, which also appear to show depth variability with amplitude $4350 \pm 1140$ ppm ($3.8 \sigma$), we realized that the transit depths had unusually large uncertainties. We attribute this to the large impact parameter; for values of $b$ close to unity, minute changes in $b$ can yield large swings in the transit depth. The most recent value of $b$ in the literature comes from \citet{Patel_2022}, which found $b= 0.846^{+0.030}_{-0.025}$. Our \textit{TESS} sector-level fits yield similar results. Thus, to improve the precision on the depths, we conducted a second round of fitting, this time fixing the impact parameter to $0.846$. This yielded noticeably better fits, with the median transit depth being $29200 \pm 1900$ ppm.

\begin{figure*}[ht!]
    \centering
    \includegraphics[width=\textwidth]{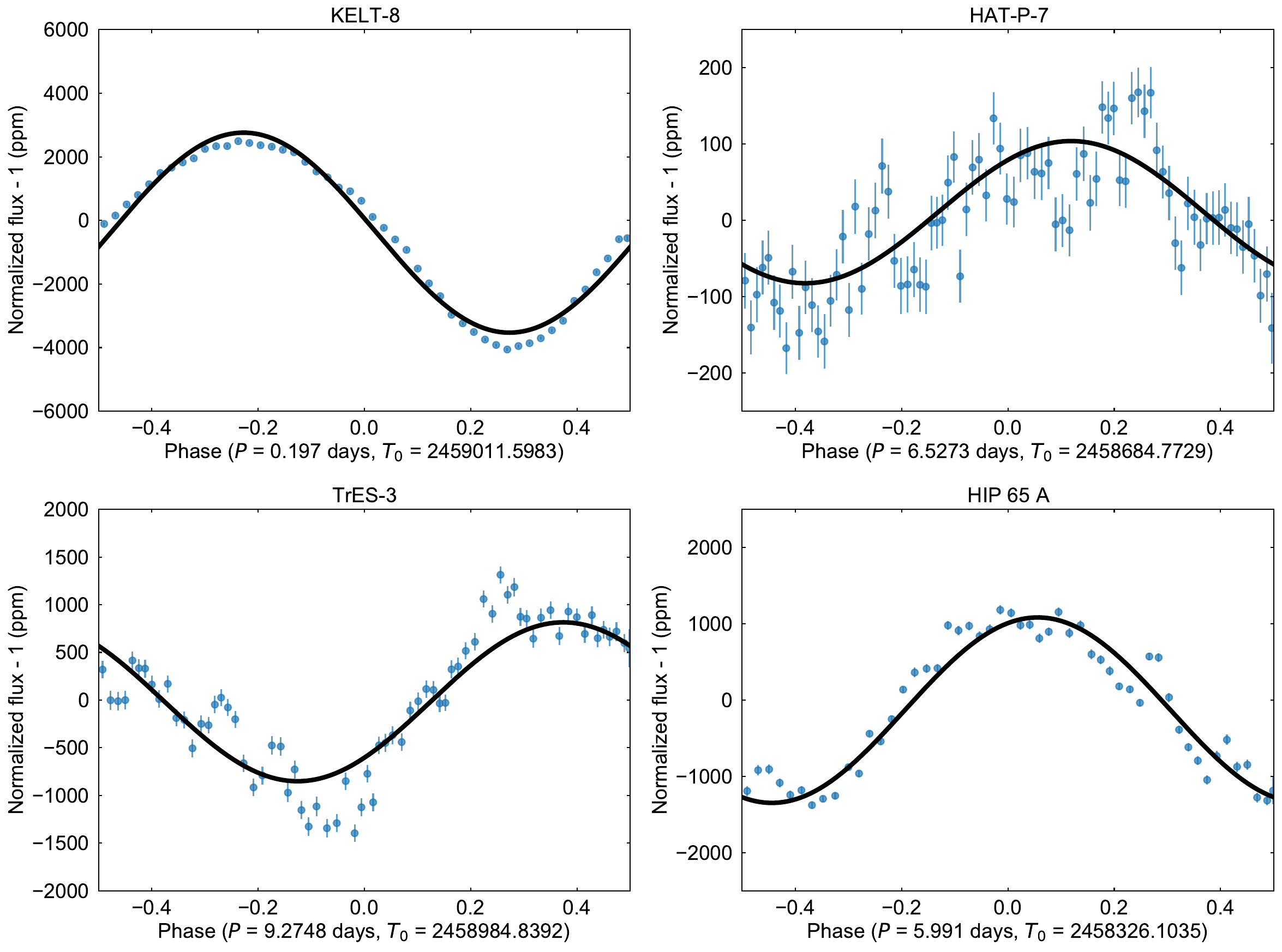}
    \caption{Phased out-of-transit lightcurve for KELT-8b (top left), HAT-P-7b (top right), TrES-3b (bottom left) and HIP 65Ab (bottom right) revealing periodic photometric activity. All lightcurves binned to 1,000 points. Best-fit sinusoidal models are overplotted in black.}
    \label{fig:oot}
\end{figure*}

Consequently, with this newer fit, the previously detected transit depth variability is no longer observed, and the periodogram of the new depths no longer has statistically significant peaks. When phased to the peak of $P = 4.2534$ days found from the first round of fitting (see Figure \ref{fig:periodograms}), the amplitude in the depths is reduced to $778 \pm 470$ ppm, only $1.7 \sigma$ from 0. The transit depths are plotted in Figure \ref{fig:phased} and do not seem to show obvious variability. Thus, we conclude that the observed peak in the periodogram likely results from the shape of the grazing transit casting uncertainty into the transit depths, creating the illusion of transit depth variability. We find photometric variability for this target to be strongest at $P = 9.3$ days with amplitude $833 \pm 17$ ppm (see Figure \ref{fig:oot}), but we do not believe this to be related to the initially observed transit depth variability. 

\subsubsection{HIP 65Ab} \label{subsubsec:hip-65ab}

HIP 65Ab \citep{Nielsen_2020}, also known as TOI-129.01 and TIC 201248411.01, was observed by \textit{TESS} in Sectors 1, 2, 28, and 29. This is a $3.2\ \text{M}_{\text{Jup}}$ planet on a 0.98-day grazing orbit. The original discovery paper finds $b=1.169^{+0.095}_{-0.077}$, along with a highly uncertain depth of $82000 ^{+52000}_{-40000}$ ppm. Our first round of fitting obtains a median depth of $10700 \pm 4600$ ppm, and phasing at the $P=26.8947$ day peak found in the periodogram, we appear to detect transit depth variability with amplitude $1600 \pm 370$ ppm ($4.3 \sigma$). However, we note that the depth is discrepant from the literature value by nearly a factor of 8. We again attribute this to the large impact parameter, which is even more extreme than TrES-3b and nominally above 1. Thus we again conduct a second round of fitting, in which we fix $b=1.169$ in our priors. This yields a revised median transit depth of $83100 \pm 2800$ ppm, with much better agreement with \citet{Nielsen_2020}. Figure \ref{fig:phased} shows the results from this fit. 

We observe that this case is similar to that of TrES-3b: upon fixing the impact parameter, the previously observed transit depth variability no longer exists. The transit depths are all consistent to well within $1 \sigma$, and the amplitude of the best-fit sinusoid is now greatly reduced to $620 \pm 1550$ ppm ($0.4 \sigma$). This is much smaller than the original detection. Although we search for and find photometric variability with period 6.0 days and amplitude $1215 \pm 6$ ppm, this does not give us further insight on the transit depth variations for this system. We again conclude that the initial transit depth variability is likely an artifact resulting from the planet's grazing orbit. 

\subsubsection{WASP-121b} \label{subsec:wasp121}

WASP-121b is a target of particular interest which has had previous claims of variability in the planet's transmission spectrum \citep{10.1093/mnras/stab797}. More specifically, spectra collected in January 2017 and October 2016 seem to show differences of $\sim 2 \sigma$ in the transit depth at wavelengths shorter than 650 nm. Here we present a brief analysis of this target. 

\begin{figure}[ht!]
    \centering
    \includegraphics[width=\columnwidth]{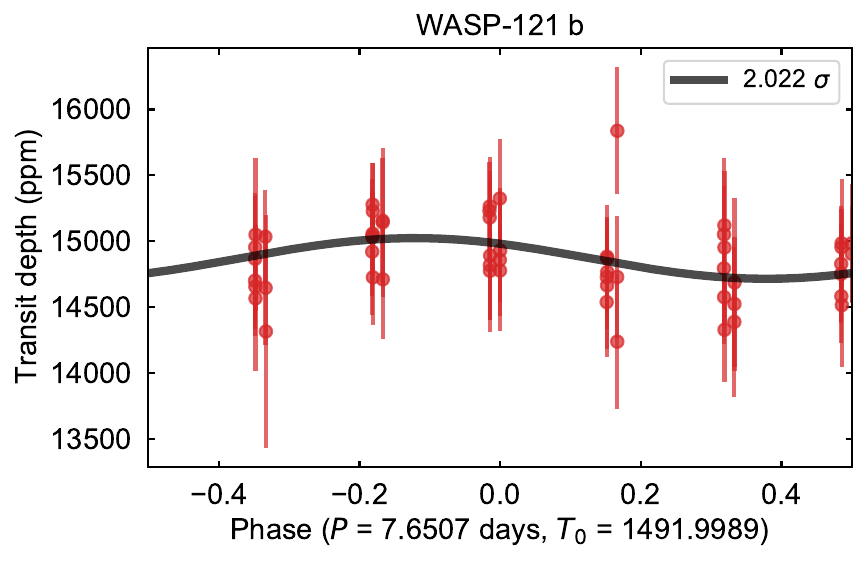}
    \caption{Phased transit depths over time for WASP-121b. Best-fit sinusoidal model overplotted in black. }
    \label{fig:121transits}
\end{figure}

We measure WASP-121b's median transit depth in the \textit{TESS} band to be $14870 \pm 290$ ppm, consistent to $2 \sigma$ with $15510 \pm 120$ ppm from \citet{10.1093/mnras/stw522}. The transit depths seem to vary with amplitude $154 \pm 76$ ppm ($2.0 \sigma$) at $P=7.6507$ days, as shown in Figure \ref{fig:121transits}, but this does not reach our $3 \sigma$ threshold). The periodogram also does not display peaks with false-alarm probabilities of lower than 0.1\%. 

\begin{figure}[ht!]
    \centering
    \includegraphics[width=\columnwidth]{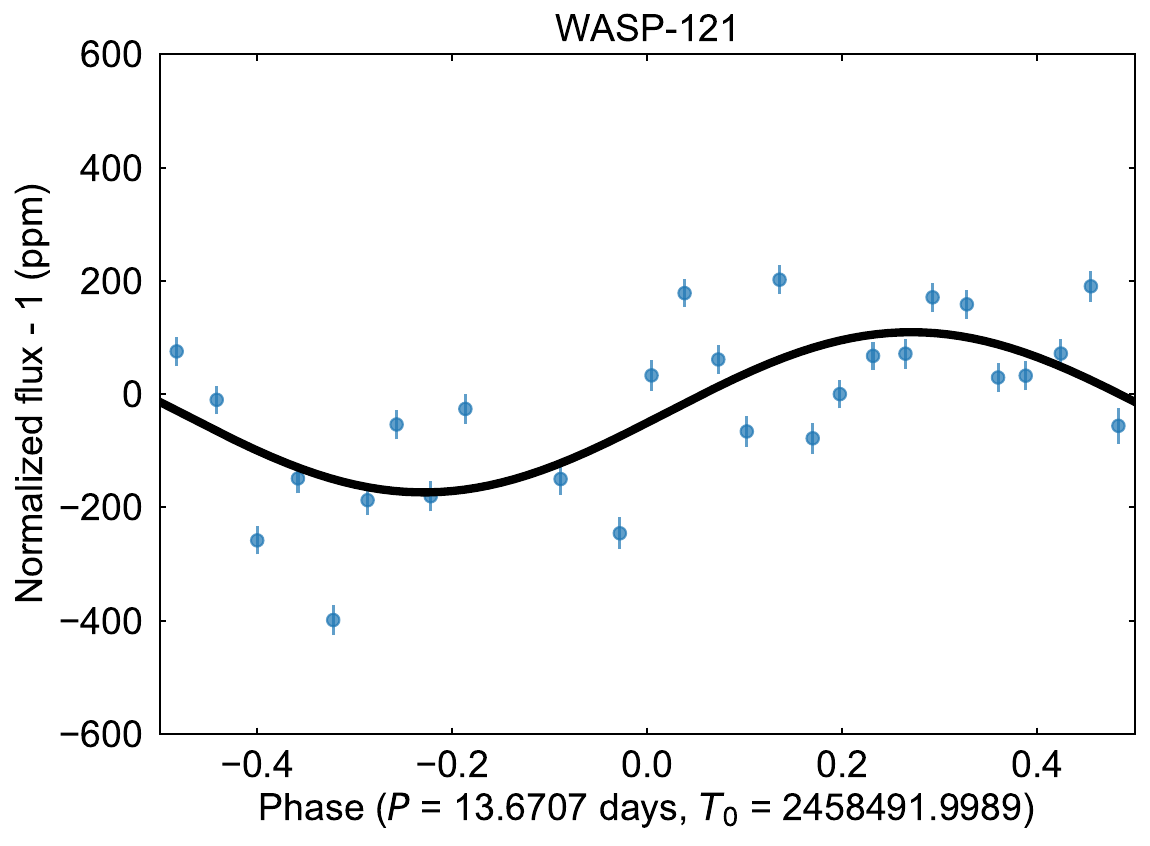}
    \caption{Phased out-of-transit lightcurve for WASP-121b, binned to 2,000 points. Best-fit sinusoidal model overplotted in black. }
    \label{fig:121oot}
\end{figure}

We detect photometric variability most strongly at a period of $P = 13.67$ days, with amplitude $141 \pm 7$ ppm (see Figure \ref{fig:121oot}). This is comparable to the variations in the transit depths. Thus, even if the transit depth variability were significant at $>3 \sigma$, we are still unable to infer planet atmospheric variability, as we cannot rule out stellar activity as an alternate explanation. Furthermore, there are a few nearby sources within $1-2$ \textit{TESS} pixels of the target star, which could also result in false-positive scenarios. 

We note, however, that \textit{TESS} has a wide, red bandpass covering $600$ to $1000$ nm, while \citet{10.1093/mnras/stab797} found that time variability for WASP-121b is most apparent for wavelengths shorter than $650$ nm. This comprises only a small portion of the \textit{TESS} bandpass (the range from $600$ to $650$ nm). In addition, \textit{TESS}'s wide bandpass could possibly dilute any true signal. Thus, we believe that while we do not detect significant transit depth variability for this target, our results are nonetheless in agreement with those of \citet{10.1093/mnras/stab797}. 

Despite this non-detection, the case of WASP-121b provides an interesting example. Because the opacity of the spectra of species such as TiO is highest in longer wavelengths around $600-1000$ nm, which coincides with the wavelength range of \textit{TESS}, targets which are observed by \textit{TESS} to exhibit transit depth variations may be more likely to experience atmospheric variability as compared to stellar activity.

\begin{figure*}
    \centering
    \includegraphics[width=\textwidth]{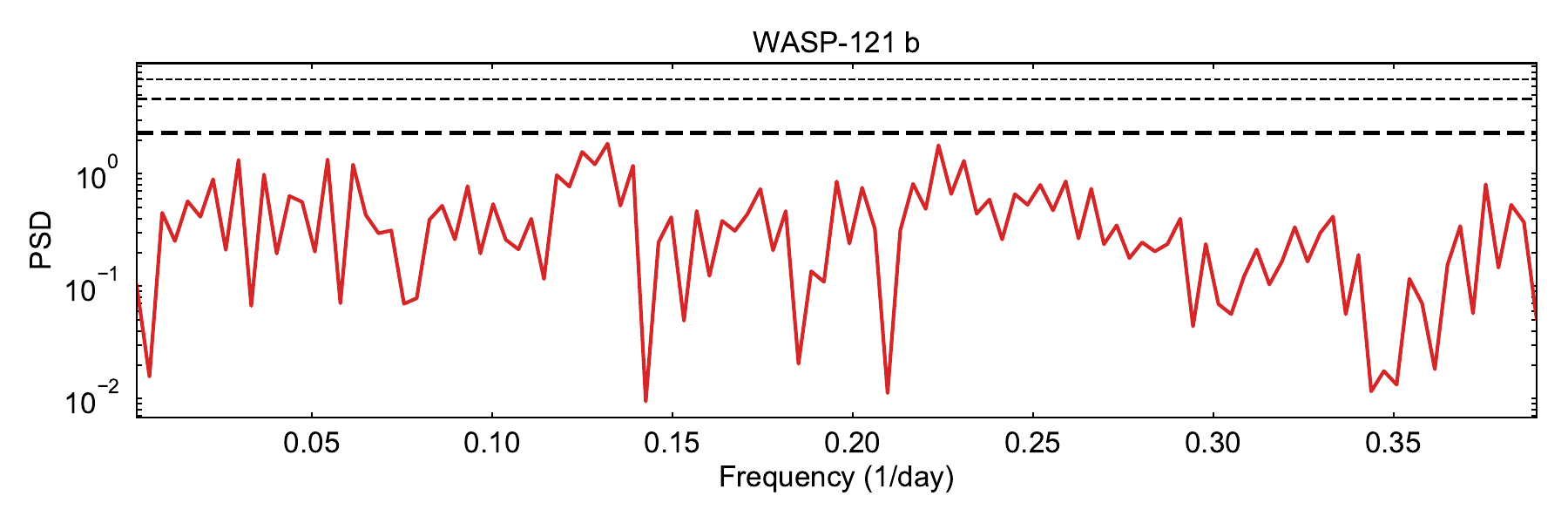}
    \caption{Periodogram for WASP-121b with dashed lines corresponding to 10\%, 1\%, and 0.1\% false-alarm probabilities. There are no peaks with false-alarm probabilities of less than 10\%.}
    \label{fig:121periodogram}
\end{figure*}

\subsection{Nested Ratio} \label{subsec:nestedresults}

Our final analysis is a search for population-level trends for transit depth variability. We hypothesize that transit depth variability may be dependent upon the strength of the Transit Light Source effect \citep{Rackham_2018}, which is in turn related to the properties of the host star. For instance, \citet{Rackham_2018} finds that this effect is very pronounced for M dwarfs, while the present state of knowledge regarding this effect for other stellar types is uncertain. To shed light on the extent to which the Transit Light Source effect may depend on stellar type, we search for trends amongst transit depth variability against stellar effective temperature and host star mass. We calculate the nested ratio, our method of normalizing transit depth variations consistently across all planets, using the steps described in Section \ref{subsec:stellar} and compare them against these two stellar properties. 

For each target, all of the transits that are injected have a constant transit depth (see Section \ref{subsubsec:simulate}). Thus, the F-ratio for a simulated set of transits is an attempt to measure the portion of the F-ratio that is solely due to the non-Gaussian noise in the lightcurve, with transit depth variability being taken out of the question. Therefore, a nested ratio greater than 1 suggests variability that persists even when accounting for noise; additionally, a larger nested ratio suggests greater true variation. We note here that such true variability may have two origins: the planet, such as its atmosphere, or the Transit Light Source effect. Thus, to focus on the latter, we exclude the four planets identified in Section \ref{subsec:periodogram_results} to possibly exhibit planetary-induced variability from this population analysis. 

Due to our simulation and selection processes (see Section \ref{subsubsec:simulate}), targets which did not meet the minimum threshold of $9$ successful injections were excluded from this analysis. Furthermore, the nature of our criteria biases against targets observed in multiple sectors, on average excluding multi-sector targets at a higher rate. However, this is not expected to bias our results. In total, $208$ of our $330$ original targets were included in this subset of the population-level analysis. The nested ratios are also listed in Table \ref{tab:mainresults}; note that some entries are missing due to the calculation only being performed on a subset of our sample. 

\subsubsection{Stellar Effective Temperature} \label{subsubsec:teff}

\begin{figure}[ht!]
    \centering
    \includegraphics[width=\columnwidth]{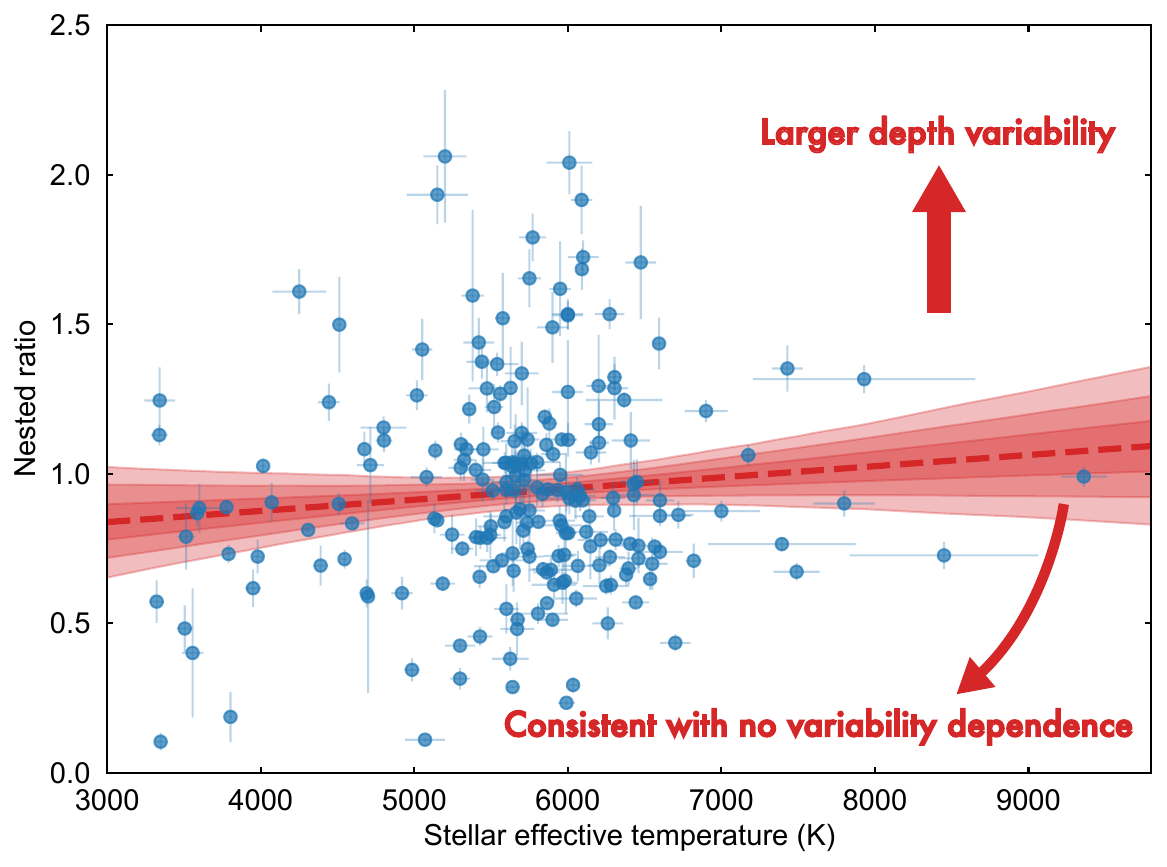}
    \caption{Nested ratio against stellar effective temperature. The best-fit linear model has slope $(3.7 \pm 2.1) \times 10^{-5}$, consistent with 0. The contours represent the 68\%, 95\%, and 99.7\% confidence intervals.}
    \label{fig:nrteff}
\end{figure}

Our first measure of stellar type is stellar effective temperature, which we obtain from various works in the literature through the \texttt{exoctk} package. Figure \ref{fig:nrteff} shows the nested ratio plotted against $T_{\text{eff}}$. Referring back to our speculation based on the results of \citet{Morales2008}, one might expect larger transit depth variability for planets orbiting cooler hosts. However, there does not seem to be a trend of higher nested ratios for cooler stars; in fact, if we ignore the errorbars, there seems to be an absence of high nested ratios ($N > 1$) for planets around stars below $\sim 5000$ K, thus suggesting that nested ratios are lower for planets around stars below $\sim 5000$ K. We note that $\le$ 5000 K corresponds to K-dwarfs, which have been found to exhibit less stellar variability and are good host stars in the search for extraterrestrial life \citep{Cuntz_2016, Lillo-Box}. 

However, this feature is not found to be statistically significant. We observe that, for stars with $T_{\text{eff}} < 5000$ K, there are 10 which have planets with $N > 1.0$ and 22 with $N < 1.0$ (69\% below 1.0), whereas for stars hotter than 5000 K we have 73 and 103 counts respectively (59\% below 1.0). A Student's t-test of the null hypothesis that these two populations have the same proportion of targets with $N < 1.0$ results in a P-value of $0.13$, which does not reach the standard $\leq 0.05$ significance threshold. Furthermore, a weighted average calculation of the nested ratios for each population reveals $N(T_{\text{eff}} < 5000 \text{K}) = 0.618 \pm 0.046$ and $N(T_{\text{eff}} > 5000 \text{K}) = 0.497 \pm 0.011$, values which are consistent at $2.5 \sigma$. Finally, we conduct an MCMC fit with bootstrapping to the data by randomly choosing two-thirds of the planets, resampling each data point using its errorbars, and performing a linear least-squares fit. We repeat this process $10^{5}$ times, and obtain slope $(3.7 \pm 2.1) \times 10^{-5}$ and vertical axis intercept (projected to $3000$ K) $0.839 \pm 0.060$, which are consistent with $0$ and $1$ to within $1.8$ and $2.7 \sigma$ respectively (see Figure \ref{fig:nrteff}). This suggests that the data is in agreement with no excess variation at the encompassed stellar temperatures. From the above tests, we conclude that our current data and analysis are not indicative of a statistically significant correlation. 

\subsubsection{Host Star Mass} \label{subsubsec:solarmass}

\begin{figure}[ht!]
    \centering
    \includegraphics[width=\columnwidth]{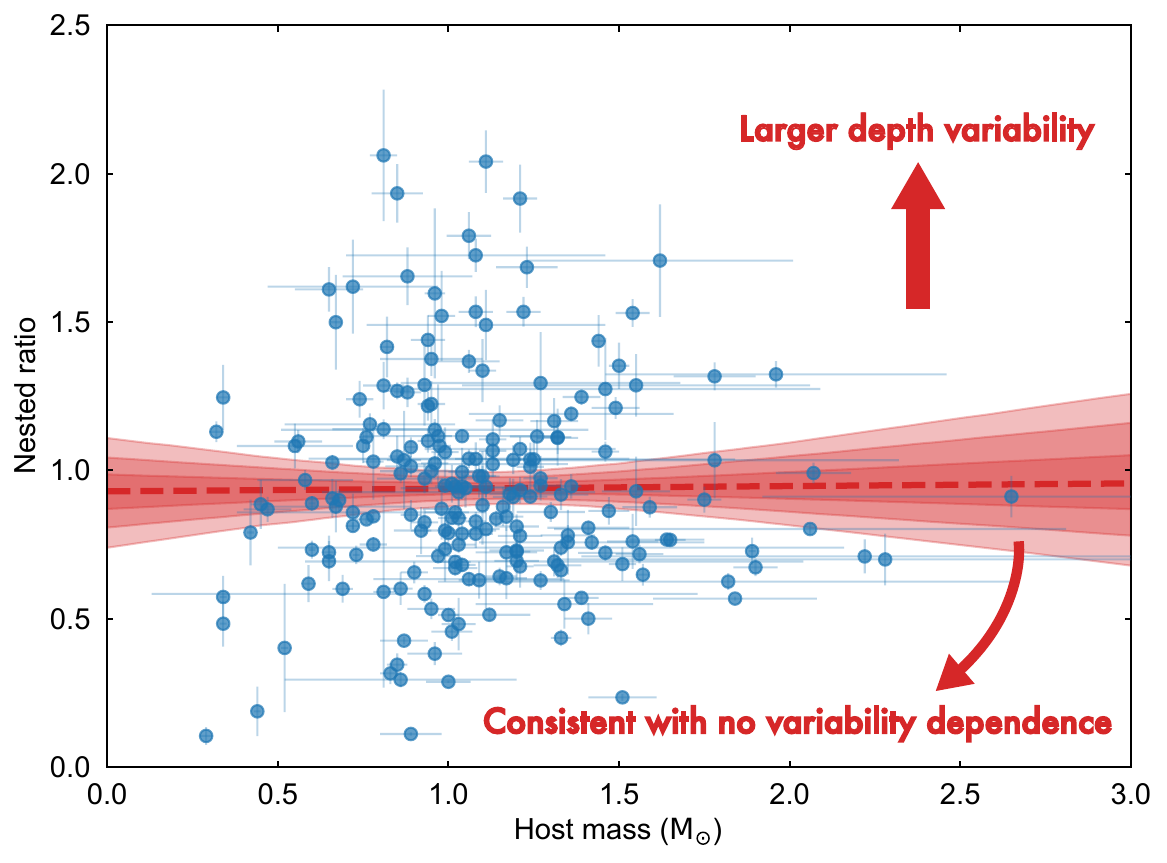}
    \caption{Same as Figure \ref{fig:nrteff}, but for nested ratio versus host star mass. The best-fit linear model has slope $(0.89 \pm 4.93) \times 10^{-2}$, again consistent with 0.}
    \label{fig:nrsolar}
\end{figure}

Our second measure of stellar type is the mass of the host star. We also obtain these through the \texttt{exoctk} package, and its relation with the nested ratio is shown in Figure \ref{fig:nrsolar}. We do not notice any obvious correlation in the data, and thus do not perform a Student's t-test and weighted average calculation. Using the same MCMC and bootstrapping linear fit as in Section \ref{subsubsec:teff} yields $(0.89 \pm 4.93) \times 10^{-2}$ for the slope and $0.929 \pm 0.059$ for the intercept, which are again consistent with $0$ and $1$ to within $0.2$ and $1.2 \sigma$. From the data we are unable to identify a significant dependence between the nested ratio and host mass. 

\section{Discussion and Conclusion} \label{sec:conclusion}

In this work we perform a blind search of transit depth variability on a set of 330 known planets observed by \textit{TESS}. We fit 2-minute PDCSAP \textit{TESS} data for all targets using \texttt{juliet}, and use out-of-transit data to inform our in-transit fits to minimize the uncertainty in our fits while keeping all transit parameters (especially transit depths) independent. For each target we calculate the standard deviation of its depths, forming an atlas of transit depth variability which can serve as reference for future depth variability studies.

Next we identify individual targets with significant signs of transit depth variability. We conduct an automated periodogram search for all targets, with further vetting conducted with a significance calculation. In general, we find that the majority of targets do not display significant variability above what is expected from noise. However, for the 4 targets KELT-8b, HAT-P-7b, TrES-3b, and HIP 65Ab -- which form roughly $1$\% of our entire sample -- we find signals that meet our selection criteria. 

For KELT-8b, we attribute the observed transit depth variability to a nearby variable star. The transit depth variability analyses for TrES-3b and HIP 65Ab are interesting as both have relatively high ($>0.8$) impact parameters, resulting in highly unstable depths when fitting for single transits. We demonstrate that keeping the impact parameter fixed in our fits stabilizes the transit depths and allows us to accurately characterize any variability. In both cases, we find that upon fixing the impact parameter, depth variability is no longer observed. This leads us to believe that the initial detection is most likely an artifact of the planets' grazing orbits. 

The only target we deem worthy of additional follow-up and analysis is HAT-P-7b. This target provides a special case in that it has also been observed by \textit{Kepler}, and its first claim of atmospheric variability had made use of \textit{Kepler} data \citep{Armstrong_2016}. It may be informative to conduct a similar measurement of transit depth variations using \textit{Kepler} data in addition to \textit{TESS}, as the combined data could provide more robust constraints on the system's observed depth variability than \textit{TESS} alone. Transit depth variability that is found to be consistent across instruments may also further demonstrate its authenticity and warrant additional follow-up such as transit spectroscopy. 

Finally, motivated by the fact that transit depth variability can be caused by stellar activity through the Transit Light Source effect \citep{Rackham_2018}, we search for population-level trends of transit depth variability against stellar type, which we characterize using stellar effective temperature and host star mass. We carry out this analysis on a subset of $208$ targets, for which we compare true transits to simulated sets of transits. This forms a nested ratio statistic that normalizes depth variability across targets. Through a variety of statistical tests, we are unable to identify any significant correlations between the nested ratio and host star properties. For individual targets with follow-up by transit spectroscopy, our atlas of transit depth variations can also serve as a guide as for how much transit depth variability could be attributed to stellar activity in the \textit{TESS} bandpass of $600-1000$ nm. 

Our work highlights the richness of datasets such as \textit{TESS}' to perform detailed analyses of known exoplanet systems, and how it can allow for the detection of relatively rare phenomena such as transit depth variations, which was the focus of study of the present work. With this data-driven view of transiting exoplanets in mind, it is also interesting to note that the Nancy Grace Roman Space Telescope, due to launch by 2027, is expected to yield $\sim 100,000$ transiting planets through the photometric monitoring of 100 million stars in its microlensing survey \citep{Montet_2017, wilson2023transiting}. With $\sim 1\%$ of targets having detectable transit depth variations in our sample, by extrapolation we could expect of order $\sim 1,000$ transiting planets discovered by \textit{Roman} to display transit depth variability at a significant level. Such a large sample could, in turn, allow us to explore in detail the cause of those variations through population level studies --- an endeavor for which our current sample is much too small. 

\section*{Acknowledgments} \label{sec:acknowledgments}

We thank the anonymous referee for their comments and suggestions, which significantly improved our results and their presentation. This work includes data collected by the TESS mission, obtained from the MAST data archive at the Space Telescope Science Institute (STScI). Funding for the TESS mission is provided by the NASA Explorer Program. STScI is operated by the Association of Universities for Research in Astronomy, Inc., under NASA contract NAS 5–26555. The TESS data described in this work may be obtained from the MAST archive at \dataset[doi.org/10.17909/t9-nmc8-f686]{https://dx.doi.org/10.17909/t9-nmc8-f686} (MAST Team \citeyear{https://doi.org/10.17909/t9-nmc8-f686}). This research has made use of the NASA Exoplanet Archive, which is operated by the California Institute of Technology, under contract with the National Aeronautics and Space Administration under the Exoplanet Exploration Program. This research has made use of the Exoplanet Follow-up Observation Program (ExoFOP; DOI: 10.26134/ExoFOP5) website, which is operated by the California Institute of Technology, under contract with the National Aeronautics and Space Administration under the Exoplanet Exploration Program. This research has made use of the open source Python package \texttt{exoctk}, the Exoplanet Characterization Toolkit \citep{matthew_bourque_2021_4556063}. 

\vspace{5mm}
\facilities{\textit{TESS}, Exoplanet Archive}

\software{\texttt{juliet} \citep{10.1093/mnras/stz2688}, \texttt{astropy} \citep{astropy:2013, astropy:2018, astropy:2022}, \texttt{dynesty} \citep{10.1093/mnras/staa278}}

\appendix

\section{Table of main results}

\setcounter{table}{0}
\renewcommand{\thetable}{A\arabic{table}}

\begin{table*}[ht!]
    \centering
    \caption{Results for all planets in our sample, with first 20 rows shown. The full table is available in the online version of this article.}  \label{tab:mainresults}
    \begin{tabular}{lcccccc}
        \hline \hline
        Planet Name & \multicolumn{1}{p{2.5cm}}{\centering Transit \\ Depth (ppm)} & \multicolumn{1}{p{2.5cm}}{\centering Transit Depth \\ Uncertainty$^{\text{a}}$ (ppm)} & \multicolumn{1}{p{2.5cm}}{\centering Absolute \\ Variability$^{\text{b}}$ \\ (ppm)} & Nested Ratio$^{\text{c}}$ & \multicolumn{1}{p{2.5cm}}{\centering Nested Ratio \\ Uncertainty} & \multicolumn{1}{p{2.0cm}}{\centering Number \\ of Transits} \\ \hline
        55 Cnc e    &$347$    &$58$    &$33$    &$\ldots$    &$\ldots$    &$88$ \\
        CoRoT-1 b    &$18976$    &$2012$    &$2033$    &$0.997$    &$0.043$    &$29$ \\
        CoRoT-2 b    &$28243$    &$1369$    &$298$    &$0.382$    &$0.039$    &$7$ \\
        CoRoT-18 b    &$27520$    &$6769$    &$4100$    &$1.375$    &$0.057$    &$8$ \\
        CoRoT-19 b    &$6009$    &$1339$    &$1346$    &$1.916$    &$0.115$    &$6$ \\
        DS Tuc A b    &$8313$    &$1209$    &$1057$    &$0.456$    &$0.031$    &$5$ \\
        G 9-40 b    &$3786$    &$1132$    &$307$    &$0.105$    &$0.030$    &$7$ \\
        GJ 357 b    &$1091$    &$178$    &$51$    &$0.483$    &$0.078$    &$7$ \\
        GJ 436 b    &$6595$    &$766$    &$643$    &$0.869$    &$0.042$    &$14$ \\
        GJ 486 b    &$1408$    &$337$    &$178$    &$1.130$    &$0.035$    &$15$ \\
        GJ 3090 b    &$1766$    &$363$    &$167$    &$0.401$    &$0.216$    &$7$ \\
        HAT-P-1 b    &$8521$    &$460$    &$341$    &$0.642$    &$0.030$    &$10$ \\
        HAT-P-2 b    &$4783$    &$152$    &$132$    &$0.663$    &$0.036$    &$16$ \\
        HAT-P-3 b    &$11221$    &$929$    &$512$    &$0.633$    &$0.025$    &$20$ \\
        HAT-P-4 b    &$7310$    &$326$    &$320$    &$1.097$    &$0.039$    &$15$ \\
        HAT-P-5 b    &$12714$    &$698$    &$586$    &$1.115$    &$0.019$    &$33$ \\
        HAT-P-7 b    &$5910$    &$292$    &$253$    &$\ldots$    &$\ldots$    &$67$ \\
        HAT-P-8 b    &$8474$    &$232$    &$210$    &$1.294$    &$0.172$    &$8$ \\
        HAT-P-9 b    &$11493$    &$755$    &$346$    &$\ldots$    &$\ldots$    &$12$ \\
        HAT-P-11 b    &$3587$    &$199$    &$153$    &$\ldots$    &$\ldots$    &$29$ \\
        \hline
    \end{tabular}

    \begin{itemize}
        \item[] $^{\text{a}}$ Transit depth uncertainty refers to the median value across individual transits for each target
        \item[] $^{\text{b}}$ Absolute variability refers to the standard deviation of the individual transit depths
        \item[] $^{\text{c}}$ Some entries are unavailable because we only performed the nested ratio analysis on a subset of our sample
    \end{itemize}
\end{table*}

\end{document}